\documentclass[runningheads]{llncs}

\usepackage{amssymb}
\usepackage{graphicx}
\usepackage{times}

\usepackage{url}

\newcommand{\ket}[1]{|#1\rangle}
\newcommand{\bra}[1]{\langle #1|}

\newcommand{\C}{\mathbb{C}}
\newcommand{\Z}{\mathbb{Z}}

\newtheorem{fact}{Fact}

\newcommand{\ra}{\rangle}
\newcommand{\la}{\langle}

\newcommand{\be}{\begin{equation}}
\newcommand{\ee}{\end{equation}}
\newcommand{\ber}{\begin{eqnarray}}
\newcommand{\eer}{\end{eqnarray}}

\begin{document}

\mainmatter  

\title{An Efficient Quantum Algorithm for the Hidden\\ Subgroup Problem over
Weyl{-}Heisenberg Groups$\!$}

\titlerunning{HSP over Weyl-Heisenberg Groups}

%
%
\author{Hari Krovi \and Martin R{\"o}tteler}


\institute{
NEC Laboratories America\\
4 Independence Way, Suite 200\\
Princeton, NJ 08540, U.S.A.\\
\texttt{$\{$krovi,mroetteler\}@nec-labs.com}
}
%

\maketitle

\begin{abstract}
  Many exponential speedups that have been achieved in quantum
  computing are obtained via hidden subgroup problems (HSPs). We show
  that the HSP over Weyl-Heisenberg groups can be solved efficiently
  on a quantum computer. These groups are well-known in physics and
  play an important role in the theory of quantum error-correcting
  codes. Our algorithm is based on non-commutative Fourier analysis of
  coset states which are quantum states that arise from a given
  black-box function. We use Clebsch-Gordan decompositions to combine
  and reduce tensor products of irreducible representations.
  Furthermore, we use a new technique of changing labels of 
  irreducible representations to obtain low-dimensional irreducible
  representations in the decomposition process.  A feature of the
  presented algorithm is that in each iteration of the algorithm the
  quantum computer operates on two coset states simultaneously.  This
  is an improvement over the previously best known quantum algorithm
  for these groups which required four coset states.
\end{abstract}


\section{Introduction}

Exponential speedups in quantum computing have hitherto been shown for
only a few classes of problems, most notably for problems that ask to
extract hidden features of certain algebraic structures. Examples for
this are hidden shift problems \cite{vDHI:2003}, hidden non-linear
structures \cite{CSV:2007}, and hidden subgroup problems (HSPs). The
latter class of hidden subgroup problems has been studied quite
extensively over the past decade. There are some successes such as the
efficient solution of the HSP for any abelian group
\cite{Shor:97,Kitaev:97,BH:97,ME:98}, including factoring and discrete
log as well as Pell's equation \cite{Hallgren:2002}, and efficient
solutions for some non-abelian groups \cite{FIMSS:2003,BCvD:2005}.
Furthermore, there are some partial successes for some non-abelian
groups such as the dihedral groups \cite{Regev:2004,Kuperberg:2005}
and the affine groups \cite{MRRS:2004}. Finally, it has been
established that for some groups, including the symmetric group which
is connected to the graph isomorphism problem, a straightforward
approach requires a rather expensive quantum processing in the sense
that entangling operations on a large number of quantum systems would
be required \cite{HMRRS:2006}. What makes matters worse, there are
currently no techniques, or even promising candidates for techniques,
to implement these highly entangling operations.

The present paper deals with the hidden subgroup problem for a class
of non-abelian groups that---in a precise mathematical sense that will
be explained below---is not too far away from the abelian case, but at
the same time has some distinct non-abelian features that make the HSP
over these groups challenging and interesting.

The hidden subgroup problem is defined as follows: we are given a
function $f: G \rightarrow S$ from a group $G$ to a set $S$, with the
additional promise that $f$ takes constant and distinct values on the
left cosets $gH$, where $g \in G$, of a subgroup $H\leq G$. The task
is to find a generating system of $H$.  The function $f$ is given as a
black-box, i.\,e., it can only be accessed through queries and in
particular whose structure cannot be further studied. The input size
to the problem is $\log{|G|}$ and for a quantum algorithm solving the
HSP to be efficient means to have a running time that is
$poly(\log{|G|})$ in the number of quantum operations as well as in
the number of classical operations.

We will focus on a particular approach to the HSP which proved to be
successful in the past, namely the so-called {\em standard method},
see \cite{GSVV:2004}. Here the function $f$ is used in a special way,
namely it is used to generate {\em coset states} which are states of
the form $1/\sqrt{|H|} \sum_{h \in H} \ket{gh}$ for random $g \in G$.
The task then becomes to extract a generating system of $H$ from a
polynomial number of coset states (for random values of $g$). A basic
question about coset states is how much information about $H$ they
indeed convey and how this information can be extracted from suitable
measurements.\footnote{Recall that the most general way to extract
  classical information from quantum states is given by means of
  positive operator valued measures (POVMs) \cite{Nielsen}.} A fixed
POVM ${\cal M}$ operates on a fixed number $k$ of coset states at once
and if $k \geq 2$ and ${\cal M}$ does not decompose into measurements
of single copies, we say that the POVM is an entangled measurement. As
in \cite{HMRRS:2006}, we call the parameter $k$ the ``jointness'' of
the measurement. It is known that information-theoretically for any
group $G$ jointness $k=O(\log{|G|})$ is sufficient \cite{EHK:2004}.
While the true magnitude of the required $k$ can be significantly
smaller (abelian groups serve as examples for which $k=1$), there are
cases for which indeed a high order of $k = \Theta(\log{|G|})$ is
sufficient {\em and} necessary.  Examples for such groups are the
symmetric groups \cite{HMRRS:2006}.  However, on the more positive
side, it is known that some groups require only a small, sometimes
even only constant, amount of jointness. Examples are the Heisenberg
groups of order $p^3$ for a prime $p$ for which $k=2$ is sufficient
\cite{BCvD:2005,Bacon:2008a}.  In earlier work \cite{ISS:2007}, it has
been shown that for the Weyl-Heisenberg groups order $p^{2n+1}$, $k=4$
is sufficient \cite{ISS:2007}.

The goal of this paper is to show that in the latter case the
jointness can be improved. We give a quantum algorithm which is
efficient in the input size (given by $\log{p}$ and $n$) and which
only requires a jointness of $k=2$.

{\bf Our results and related work:} The family of groups we consider
in the present paper are well-known in quantum information processing
under the name of generalized Pauli groups or Weyl-Heisenberg groups
\cite{Nielsen}.  Their importance in quantum computing stems from the
fact that they are used to define stabilizer codes, the class of codes
most widely used for the construction of quantum error-correcting
codes \cite{CRSS96,Gottesman96,CRSS98}.

In a more group-theoretical context, the Weyl-Heisenberg groups are
known as extraspecial $p$-groups (actually, they constitute one of the
two families of extraspecial $p$-groups \cite{Huppert:83}). A
polynomial-time algorithm for the HSP for the extraspecial $p$-groups
was already given by Ivanyos, Sanselme, and Santha, \cite{ISS:2007}.
Our approach differs to this approach in two aspects: first, our
approach is based on Fourier sampling for the non-abelian group $G$.
Second, and more importantly, we show that the jointness $k$, i.\,e.,
the number of coset states that the algorithm has to operate jointly
on, can be reduced from $k=4$ to $k=2$. Crucial for our approach is
the fact that in the Weyl-Heisenberg group the labels of irreducible
representations can be changed. This is turn can be used to ``drive''
Clebsch-Gordan decompositions in such a way that low-dimensional
irreducible representations occur in the decomposition.

It is perhaps interesting to note that for the Weyl-Heisenberg groups
the states that arise after the measurement in the Fourier sampling
approach (also called Fourier coefficients) are typically of a very
large rank (i.\,e., exponential in the input size). Generally, large
rank usually is a good indicator of the intractability of the HSP,
such as in case of the symmetric group when $H$ is a full support
involution. Perhaps surprisingly, in the case of the Weyl-Heisenberg
group it still is possible to extract $H$ efficiently even though the
Fourier coefficients have large rank. We achieve this at the price of
operating on two coset states at the same time. This leaves open the
question whether $k=1$ is possible, i.\,e., if the hidden subgroup $H$
can be identified from measurements on single coset states. We cannot
resolve this question but believe that this will be hard. Our
reasoning is as follows. Having Fourier coefficients of large rank
implies that the random basis method \cite{RRS:2005,Sen:2006} cannot
be applied. The random basis method is a method to derive algorithms
with $k=1$ whose quantum part can be shown to be polynomial, provided
that the rank of the Fourier coefficients is constant.\footnote{This
  can be obtained by combining the random basis method \cite{Sen:2006}
  with the derandomization results of \cite{AE:2007}.}  Based on this
we therefore conjecture that any efficient quantum algorithm for the
extraspecial groups will require jointness of $k\geq 2$.

Finally, we mention that a similar method to combine the two registers
in each run of the algorithm has been used by Bacon \cite{Bacon:2008a}
to solve the HSP in the Heisenberg groups of order $p^3$. The method
uses a Clebsch-Gordan transform which is a unitary transform that
decomposes the tensor product of two irreducible representations
\cite{Serre:77} into its constituents. The main difference between the
Heisenberg group and the Weyl-Heisenberg groups is that the Fourier
coefficients are no longer pure states and are of possibly high rank.

{\bf Organization of the paper:} In Section \ref{sec:esg} we review
the Weyl-Heisenberg group and its subgroup structure. The Fourier
sampling approach and the so-called standard algorithm are reviewed in
Section \ref{sec:FourierApproach}. In Section \ref{sec:irreps} we
provide necessary facts about the representation theory that will be
required in the subsequent parts. The main result of this paper is the
quantum algorithm for the efficient solution of the HSP in the
Weyl-Heisenberg groups presented in Section \ref{sec:qalgo}. Finally,
we offer conclusions in Section \ref{sec:conclusions}.

\section{The Weyl-Heisenberg groups}\label{sec:esg}

We begin by recalling some basic group-theoretic notions. Recall that
the center $Z(G)$ of a group $G$ is defined as the set of elements
which commute with every element of the group i.e., $Z(G) = \{c :
[c,g]=cgc^{-1}g^{-1}=e \;\mbox{for all}\;g \in G\}$, where $e$ is the
identity element of $G$. The derived (or commutator) subgroup $G'$ is
generated by elements of the type $[a,b]=aba^{-1}b^{-1}$, where
$a,b\in G$. The reader is invited to recall the definition of
semidirect products $G=N \rtimes H$, see for instance
\cite{Huppert:83,Serre:77}. In the following we give a definition of
the Weyl-Heisenberg groups as a semidirect product and give two
alternative ways of working with these groups.

\begin{definition} Let $p$ be a prime and let $n$ be an integer. The
  Weyl-Heisenberg group of order $p^{2n+1}$ is defined as the
  semidirect product $\mathbb{Z}_p^{n+1}\rtimes_\phi\mathbb{Z}_p^n$,
  where the action $\phi$ in the semidirect product is defined on
  $x=(x_1, \ldots, x_n) \in \Z_p^n$ as the $(n+1)\times (n+1)$ matrix
  given by \be \phi(x)= \left(\begin{array}{cccc}
      1 & \dots & 0 & 0\\
      0 & 1 & \dots & 0\\
      &\ddots &\ddots \\
      0 & \dots & 1 & 0\\
      x_1 & x_2 & \dots x_n & 1
\end{array}\right).  \ee 
\end{definition}

Any group element of $\mathbb{Z}_p^{n+1}\rtimes_\phi\mathbb{Z}_p^n$
can be written as a triple $(x,y,z)$ where $x$ and $y$ are vectors of
length $n$ whose entries are elements of $\mathbb{Z}_p$ and $z$ is in
$\mathbb{Z}_p$. To relate this triple to the semidirect product, one
can think of $(y,z)\in\mathbb{Z}_p^{n+1}$ and $x\in\mathbb{Z}_p^n$.
Then, the product of two elements in this group can be written as
\be\label{inner_prod} (x,y,z)\cdot (x',y',z')=(x+x',y+y',z+z'+x'\cdot
y), \ee where $x\cdot y=\sum_ix_iy_i$ is the dot product of two
vectors (denoted as $xy$ in the rest of the paper).

\begin{fact} \cite{Huppert:83} For any $p$ prime, and $n\geq 1$, the
  Weyl-Heisenberg group is an extraspecial $p$ group. Recall that a
  group $G$ is extraspecial if $Z(G)=G'$, the center is isomorphic to
  $\Z_p$, and $G/G'$ is a vector space.
\end{fact}

Up to isomorphism, extraspecial $p$-groups are of two types: groups of
exponent $p$ and groups of exponent $p^2$. The Weyl-Heisenberg groups
are the extraspecial $p$-groups of exponent $p$. It was shown in
\cite{ISS:2007} that an algorithm to find hidden subgroups in the
groups of exponent $p$ can be used to find hidden subgroups in groups
of exponent $p^2$.  Therefore, it is enough to solve the HSP in groups
of exponent $p$. In this paper, we present an efficient algorithm for
the HSP over groups of exponent $p$.

\paragraph{Realization via matrices over $\Z_p$:} First, we recall that the
Heisenberg group of order $p^3$ (which is the group of $3\times 3$
upper triangular matrices with ones on the main diagonal and other
entries in $\mathbb{Z}_p$) is a Weyl-Heisenberg group and can be
regarded as the semidirect product
$\mathbb{Z}_p^2\rtimes\mathbb{Z}_p$. An efficient algorithm for the
HSP over this group is given in \cite{BCvD:2005}. Elements of this
group are of the type \be \left(\begin{array}{ccc}
    1 & y & z\\
    0 & 1 & x\\
    0 & 0 & 1\\
\end{array}\right). \nonumber
\ee
The product of two such elements is
\be
\left(\begin{array}{ccc}
1 & y & z\\
0 & 1 & x\\
0 & 0 & 1\\
\end{array}\right)
\left(\begin{array}{ccc}
1 & y' & z'\\
0 & 1 & x'\\
0 & 0 & 1\\
\end{array}\right)
=
\left(\begin{array}{ccc}
1 & y+y' & z+z'+x'y\\
0 & 1 & x+x'\\
0 & 0 & 1\\
\end{array}\right)
\nonumber \ee Thus, such a matrix can be identified with a triple
$(x,y,z)$ in $\mathbb{Z}_p^2\rtimes\mathbb{Z}_p$. This matrix
representation of the Heisenberg group can be generalized for any $n$.
We can associate a triple $(x,y,z)$ where $x,y\in\mathbb{Z}_p^n$ and
$z\in\mathbb{Z}_p$ with the $(n+2)\times (n+2)$ matrix \be
\left(\begin{array}{ccccc}
    1 & y_1 & \dots & y_n & z\\
    0 & 1 & \dots & 0 & x_1\\
    \ddots & \ddots & \dots & \ddots & \ddots\\
    0 & 0 & \dots & 1 & x_n\\
    0 & 0 & \dots & 0 & 1
\end{array}\right).  \ee 

\paragraph{Realization via unitary representation:} Finally, there is
another useful way to represent the Weyl-Heisenberg group. The $n$
qu$p$it Pauli matrices form a faithful (irreducible) representation of
the Weyl-Heisenberg $p$-group.  For any $k\not=0$, we can associate
with any triple $(x,y,z)$ in
$\mathbb{Z}_p^{n+1}\rtimes\mathbb{Z}_p^n$, the following matrix: \be
\rho_k(x,y,z)=\omega_p^{kz}X^xZ_k^y, \ee where the matrix
$X=\sum_{u\in \mathbb{Z}_p^n}|u+1\ra\la u|$ is the generalized $X$
operator and the matrix $Z_k=\sum_{u\in
  \mathbb{Z}_p^n}\omega_p^k|u\ra\la u|$ is the generalized $Z$
operator, see e.\,g. \cite{Nielsen}.

\paragraph{Subgroup structure:}
In the following we will write $G$ in short for Weyl-Heisenberg
groups. Using the notation introduced above the center $Z(G)$ (or
$G'$) is the group $Z(G)=\{(0,0,z)|z\in \mathbb{Z}_p\}$ and is
isomorphic to $\mathbb{Z}_p$. As mentioned above, the quotient group
$G/G'$ is a vector space isomorphic to $\mathbb{Z}_p^{2n}$. This space
can be regarded as a {\em symplectic} space with the following inner
product: $(x,y)\cdot (x',y')=(x\cdot y'-y\cdot x')$, where $x,y,x',y'
\in \mathbb{Z}_p^n$. The quotient map is just the restriction of the
triple $(x,y,z)\in G$ to the pair $(x,y)\in \mathbb{Z}_p^{2n}$. From
Eq.~(\ref{inner_prod}), it follows that two elements commute if and
only if $xy'-yx'=0$. Denote the set of $(x,y)$ pairs occurring in $H$
as $S_H$ i.e., for each triple $(x,y,z)\in H$, we have that $(x,y)\in
S_H$ and so $|S_H|\leq |H|$. It can be easily verified that $S_H$ is a
vector space and is in fact, a subspace of $\mathbb{Z}_p^{2n}$.
Indeed, for two elements $(x,y),(x',y')\in S_H$, pick two elements
$(x,y,z),(x',y',z')\in H$ and so $(x+x',y+y',z+z'+x'y)\in H$.
Therefore, $(x+x',y+y')\in S_H$. To show that if $(x,y)\in S_H$, then
$(ax,ay)\in S_H$ for any $a\in \mathbb{Z}_p$, observe that if
$(x,y,z)\in H$, then $(x,y,z)^a=(ax,ay,az+\frac{a(a-1)}{2}xy)\in H$.
Therefore, $(ax,ay)\in S_H$ (in fact, it can be shown that $S_H\simeq
HG'/G'$, but we do not need this result.) Therefore, $H\leq G$ is
abelian if and only if $\forall (x,y), (x',y') \in S_H$, we have that
$xy'-x'y=0$. Such a space where all the elements are orthogonal to
each other is called {\em isotropic}.

Now, we make a few remarks about the conjugacy class of some subgroup
$H$.  Consider conjugating $H$ by some element of $G$, say
$g=(x',y',z')$. For any $h=(x,y,z)\in H$, we obtain
\begin{eqnarray}
g^{-1}hg\nonumber
&=&(-x',-y',-z'+x'y')(x,y,z)(x',y',z')\nonumber \\
&=&(-x',-y',-z'+x'y')(x+x',y+y',z+z'+x'y)\nonumber \\
&=& (x,y,z+x'y-xy') \in H^g.\label{conj}
\end{eqnarray}
From this we see that $S_{H^g}=S_H$. We show next that $S_H$ actually
characterizes the conjugacy class of $H$. Before proving this result
we need to determine the stabilizer of $H$. The stabilizer $H_S$ of
$H$ is defined as the set of elements of $G$ which preserve $H$ under
conjugation i.e., $H_S=\{g\in G | H^g=H\}$. From Eq.~(\ref{conj}), we
can see that $g=(x',y',z')\in H_S$ if and only if $x'y-xy'=0$ for all
$(x,y,z)\in H$. Thus, the stabilizer is a group such that
$S_{H_S}=S_H^\perp$, where $S_H^\perp$ is the orthogonal space under
the symplectic inner product defined above, i.e., $H_S=\{(x,y,z)\in G
| (x,y)\in S_H^\perp, z\in \mathbb{Z}_p\}$. In other words, it is
obtained by appending the pairs $(x,y)\in S_H^\perp$ with every
possible $z\in \mathbb{Z}_p$. Therefore, $|H_S|=|G'|\cdot
|S_H^\perp|$. Now, we can prove the following lemma.
\begin{lemma}\label{lemma:conj}
Two subgroups $H_1$ and $H_1$ are conjugate if and only if $S_{H_1}=S_{H_2}$.
\end{lemma}
\begin{proof}
  We have already seen that if $H_1$ and $H_2$ are conjugates, then
  $S_{H_1}=S_{H_2}$. To show the other direction, we use a counting
  argument ie., we show that the number of subgroups $H'$ of $G$ such
  that $S_{H'}=S_H$ is equal to the number of conjugates of $H$.
  First, assume that the dimension of the vector space $S_{H_1}$ is
  $k$. Now, the number of conjugates of $H_1$ is the index of the
  stabilizer of $H_1$. From the above result, the stabilizer has a
  size $|G'||S_{H_1}^\perp|=p\cdot p^{2n-k}$. Therefore, the index or
  the number of conjugates of $H_1$ are $p^{2n+1}/p^{2n-k-1}=p^k$.
  Now, the number of different possible subgroups $H$ such that
  $S_H=S_{H_1}$ is $p^k$ since each of the $k$ basis vectors of
  $S_{H_1}$ are generators of the subgroup and they can have any $z$
  component independent of each other i.e., there are $p$ possible
  choices of $z$ for each of the $k$ generators.
\end{proof}

The property $G'=Z(G)$ will be useful in that it will allow us to
consider only a certain class of hidden subgroups. We show next that
it is enough to consider hidden subgroups which are abelian and do not
contain $G'$. Recall that that $H$ is normal in $G$ (denoted $H\unlhd
G$) if $g^{-1}hg\in H$ for all $g\in G$ and $h\in H$.

\begin{lemma}\label{lem1}
If $G'\leq H$, then $H\unlhd G$.
\end{lemma}
\begin{proof}
  Since $G'$ is the commutator subgroup, for any $g_1,g_2\in G$, there
  exists $g'\in G'$ such that $g_1g_2=g_2g_1g'$. Now, let $h\in H$
  and $g\in G$. We have $g^{-1}hg=hg'$ for some $g'\in G'$. But since
  $G'\leq H$, $hg'=h'$, for some $h'\in H$. Therefore, $g^{-1}hg=h'$
  and hence $H\unlhd G$.
\end{proof}
\begin{lemma}
If $H$ is non-abelian, then $H\unlhd G$.
\end{lemma}
\begin{proof}
  Let $h_1,h_2\in H$ such that $h_1h_2\neq h_2h_1$. Then
  $h_1h_2=h_2h_1g'$ for some $g'\in G'$ such that $g'\neq e$, where
  $e$ is the identity element of $G$. This means that $g'\in H$. Since
  $G'$ is cyclic of prime order, it can be generated by any $g'\neq e$
  and hence, we have $G'\leq H$. Now, Lemma \ref{lem1} implies that
  $H\unlhd G$.
\end{proof}
From these two lemmas, we have only two cases to consider for the
hidden subgroup $H$: $(a)$ $H$ is abelian and does not contain $G'$
and $(b)$ $H$ is normal in $G$. It is possible to tell the cases apart
by querying the hiding function $f$ twice and checking whether $f(e)$
and $f(g')$ are equal for some $g'\neq e$ and $g'\in G'$. If they are
equal then $G'\leq H$ and $H\unlhd G$, otherwise $H$ is abelian. If
$H$ is normal, then one can use the algorithm of \cite{HRT:2003},
which is efficient if one can intersect kernels of the irreducible
representations (irreps) efficiently. For the Weyl-Heisenberg group,
the higher dimensional irreps form a faithful representation and hence
do not have a kernel. Thus, when the hidden subgroup is normal, only
one dimensional irreps occur and their kernels can be intersected
efficiently and the hidden subgroup can be found using the algorithm
of \cite{HRT:2003}. Therefore, we can consider only those hidden
subgroups which are abelian and moreover do not contain $G'$.

Now, we restrict our attention to the case of abelian $H$.
Finally, we need the following two results.
\begin{lemma}
If $H$ is an abelian subgroup which does not contain $G'$, then $|S_H|=|H|$. 
\end{lemma}
\begin{proof} Suppose that for some $(x,y)\in S_H$ there exist two
  different elements $(x,y,z_1)$ and $(x,y,z_2)$ in $H$, then by
  multiplying one with the inverse of the other we get
  $(0,0,z_1-z_2)$. Since $z_1-z_2\neq 0$, this generates $G'$, but by
  our assumption on $H$, $G'\nleq H$.  Therefore, $|S_H|=|H|$.
\end{proof}
The following theorem applies to the case when $p>2$.
\begin{lemma} Let $H$ be an abelian subgroup which does not contain
  $G'$. There exists a subgroup $H_0$ conjugate to $H$, where
  $H_0=\{(x,y,xy/2) | (x,y)\in S_H\}$.
\end{lemma}
\begin{proof}
  We can verify that $H_0$ is a subgroup by considering elements
  $(x,y,xy/2)$ and $(x',y',x'y'/2)$ in $H_0$. Their product is \ber
  (x,y,xy/2)\cdot (x',y',x'y'/2)&=&(x+x',y+y',xy/2+x'y'/2+x'y) \nonumber \\
  &=& (x+x',y+y',xy/2+x'y'/2+(x'y+xy')/2) \nonumber \\
  &=& (x+x',y+y',(x+x')(y+y')/2), \eer which is an element of $H_0$.
  Here, we have used the fact that $H$ is abelian i.e., $xy'-x'y=0,
  \forall (x,y),(x',y')\in S_H$.  Now for $H_0$, since $S_{H_0}=S_H$,
  $H_0$ is conjugate to $H$ using Lemma \ref{lemma:conj}.
\end{proof}
Note that $H_0$ can be thought of as a representative of the conjugacy
class of $H$ since it can be uniquely determined from $S_H$. The above
lemma does not apply for the case $p=2$. When $p=2$, we have that
$(x,y,z)^2=(2x,2y,2z+xy)=(0,0,xy)$. But since we assume that $G'\nleq
H$, when $p=2$ we must have that $xy=0$, $\forall (x,y,z)\in H$.

\section{Fourier sampling approach to HSP}\label{sec:FourierApproach}

We recall some basic facts about the Fourier sampling approach to the
HSP, see also \cite{GSVV:2004,HMRRS:2006}. First, we recall some basic
notions of representation theory of finite groups \cite{Serre:77} that
are required for this approach. Let $G$ be a finite group, let $\C[G]$
to denote its group algebra, and let $\hat{G}$ be the set of
irreducible representations (irreps) of $G$. We will consider two
distinguished orthonormal vector space bases for $\C[G]$, namely, the
basis given by the group elements on the one hand (denoted by
$\ket{g}$, where $g\in G$) and the basis given by normalized matrix
coefficients of the irreducible representations of $G$ on the other
hand (denoted by $\ket{\rho,i,j}$, where $\rho\in \hat{G}$, and
$i,j=1,\ldots,d_\rho$ for $d_\rho$, where $d_\rho$ denotes the
dimension of $\rho$). Now, the quantum Fourier transform over $G$,
${\rm QFT}_G$ is the following linear transformation
\cite{Beth:87,GSVV:2004}:
\begin{equation}\label{eq:QFTdef}
\ket{g} \mapsto \sum_{\rho\in \hat{G}} \sqrt{\frac{d_\rho}{|G|}}
\sum_{i, j = 1}^{d_\rho} \rho_{i j}(g) \ket{\rho, i, j}.
\end{equation}
An easy consequence of Schur's Lemma is that ${\rm QFT}_G$ is a
unitary transformation in $\C^{|G|}$, mapping from the basis of
$\ket{g}$ to the basis of $\ket{\rho,i,j}$. For a subgroup $H \leq G$
and irrep $\rho \in \hat{G}$, define $\rho(H) := \frac{1}{|H|} \sum_{h
  \in H} \rho(h)$. Again from Schur's Lemma we obtain that $\rho(H)$
is an orthogonal projection to the space of vectors that are
point-wise fixed by every $\rho(h)$, $h \in H$.

Define $r_\rho(H) := \mbox{\rm rank}(\rho(H))$; then $r_\rho(H) =
{1}/{|H|} \sum_{h \in H} \chi_\rho(h)$, where $\chi_\rho$ denotes
the character of $\rho$.  For any subset $S \leq G$ define $\ket{S} :=
{1}/{\sqrt{|S|}} \sum_{s \in S} \ket{s}$ to be the uniform
superposition over the elements of $S$.

The {\em standard method}~\cite{GSVV:2004} starts from
${1}/{\sqrt{|G|}} \sum_{g \in G} \ket{g}\ket{0}$. It then queries $f$
to get the superposition ${1}/{\sqrt{|G|}} \sum_{g \in G} \ket{g}
\ket{f(g)}$. The state becomes a mixed state given by the density
matrix $\sigma^G_H = \frac{1}{|G|} \sum_{g \in G} \ket{gH}\bra{gH}$ if
the second register is ignored.  Applying ${\rm QFT}_G$ to
$\sigma^G_H$ gives the density matrix
\[
\frac{|H|}{|G|} \bigoplus_{\rho \in \hat{G}} \bigoplus_{i=1}^{d_\rho}
\ket{\rho,i}\bra{\rho,i} \otimes \rho^\ast(H),
\]
where $\rho^\ast(H)$ operates on the space of column indices of
$\rho$. The probability distribution induced by this base change is
given by $P(\mbox{observe}\; \rho) = \frac{d_{\rho} |H|
  r_{\rho}(H)}{|G|}$. It is easy to see that measuring the rows does
not furnish any new information: indeed, the distribution on the row
indices is a uniform distribution $1/d_\rho$. The reduced state on the
space of column indices on the other hand can contain information
about $H$: after having observed an irrep $\rho$ and a row index $i$,
the state is now collapsed to $\rho^\ast(H)/r_\rho(H)$. From this
state we can try to obtain further information about $H$ via
subsequent measurements.

Finally, we mention that Fourier sampling on $k\geq 2$ registers can
be defined in a similar way.  Here one starts off with $k$ independent
copies of the coset state and applies ${\rm QFT}_G^{\otimes k}$ to it.
In the next section, we describe the representation theory of the
Weyl-Heisenberg groups. An efficient implementation of ${\rm QFT}_G$
is shown in Appendix \ref{ap:qft}.

\section{The irreducible representations}
\label{sec:irreps}

In this section, we discuss the representation theory of $G$, where
$G\cong \Z_p^{n+1} \rtimes \Z_p^n$ is a Weyl-Heisenberg group. From
the properties of being an extraspecial group, it is easy to see that
$G$ has $p^{2n}$ one dimensional irreps and $p-1$ irreps of dimension
$p^n$. The one dimensional irreps are given by \be
\chi_{a,b}(x,y,z)=\omega_p^{(ax+by)}, \ee where $\omega_p=e^{2\pi
  i/p}$ and $a,b\in \mathbb{Z}_p^n$. Note that \be
\chi_{a,b}(H)=\frac{1}{|H|}\sum_{(x,y,z)\in
  H}\omega_p^{ax+by}=\frac{1}{|S_H|}\sum_{(x,y)\in
  S_H}\omega_p^{ax+by} .  \ee Since $S_H$ is a linear space, this
expression is non-zero if and only if $a,b\in S_H^\perp$. Suppose we
perform a QFT on a coset state and measure an irrep label.
Furthermore, suppose that we obtain a one dimensional irrep (although
the probability of this is exponentially small as we show in the next
section). Then this would enable us to sample from $S_H^\perp$. If
this event of sampling one dimensional irreps would occur some $O(n)$
times, we would be able to compute a generating set of $S_H^\perp$
with constant probability. This gives us information about the
conjugacy class of $H$ and from knowing this, it is easy to see that
generators for $H$ itself can be inferred by means of solving a
suitable abelian HSP.

Thus, obtaining one dimensional irreps would be useful. Of course we
cannot assume to sample from one dimensional irreps as they have low
probability of occurring. Our strategy will be to ``manufacture'' one
dimensional irreps from combining higher-dimensional irreps. First,
recall that the $p^n$ dimensional irreps are given by \be
\rho_k(x,y,z)=\sum_{u\in \mathbb{Z}_p^n}\omega_p^{k(z+yu)}|u+x\ra\la
u| , \ee where $k\in\mathbb{Z}_p$ and $k\neq 0$.  This representation
is a faithful irrep and its character is given by $\chi_k(g)=0$ for
$g\neq e$ and $\chi_k(e)=p^n$. In particular, $\chi_k(H)=p^n/|H|$.

The probability of a high dimensional irrep occurring in Fourier
sampling is very high (we compute this in Section \ref{sec:qalgo}). We
consider the tensor product of two such high dimensional irreps. This
tensor product can be decomposed into a direct sum of irreps of the
group. A unitary base change which decomposes such a tensor product
into a direct sum of irreps is called a {\em Clebsch-Gordan}
transform, denoted by $U_{CG}$. Clebsch-Gordan transforms have been
used implicitly to bound higher moments of a random variable that
describes the probability distribution of a POVM on measuring a
Fourier coefficient.  They have also been used in \cite{Bacon:2008a}
to obtain a quantum algorithm for the HSP over Heisenberg groups of
order $p^3$, and in \cite{Bacon:2008b} for the HSP in the groups
$D_4^n$ as well as for Simon's problem. Our use of Clebsch-Gordan
transforms will be somewhat similar.

For the Weyl-Heisenberg group $G$, the irreps that occur in the
Clebsch-Gordan decomposition of the tensor product of high dimensional
irreps $\rho_k(g)\otimes\rho_l(g)$ depend on $k$ and $l$. The
Clebsch-Gordan transform for $G$ is given by 
\begin{eqnarray}
  U_{CG}:|u,v\ra\rightarrow\left\{
\begin{array}{cl}
   \sum_{w\in\mathbb{Z}_p^n}\omega_p^{\frac{l}{2}(u+v)w}|u-v,w\ra  \mbox{ for }k+l=0 \\
         |u-v,\frac{ku+lv}{k+l}\ra \mbox{ for }k+l\neq 0
\end{array} \right.
\end{eqnarray}
If $k+l\neq 0$, then only one irrep of $G$ occurs with multiplicity
$p^n$, namely \be
\rho_k(g)\otimes\rho_l(g)\stackrel{U_{CG}}{\rightarrow}
I_{p^n}\otimes\rho_{k+l}(g).  \ee If $k+l=0$, then all the one
dimensional irreps occur with multiplicity one i.e., \be
\rho_k(g)\otimes\rho_l(g)\stackrel{U_{CG}}{\rightarrow} \oplus_{a,b\in
  \mathbb{Z}_p} \chi_{a,b}(g).  \ee Note, however, that the state
obtained after Fourier sampling is not $\frac{1}{|H|}\sum_{g \in H}
\rho_k(g)\otimes\rho_l(g)$, but rather $\rho_k(H)\otimes\rho_l(H)$.
When we apply the Clebsch-Gordan transform to this state, we obtain
one dimensional irreps $\chi_{a,b}(H)$ on the diagonal.  Applying this
to $\rho_{-l}(H)\otimes\rho_l(H)$ gives us
\begin{eqnarray*}
&& \sum_{\stackrel{(x,y,z),(x',y',z')\in H}{u,v,w_1,w_2\in \mathbb{Z}_p^n}}\omega_p^{-l(yu+z)+l(y'v+z')+\frac{l}{2}((u+v)(w_1-w_2)+w_1(x+x'))\times}\\[-4ex]
&&\hspace*{3cm} |u-v+x-x',w_1\ra\la u-v,w_2|\\[2ex]
&=&\sum_{\stackrel{(x,y,z),(x',y',z')\in H}{u',w_1,w_2\in
    \mathbb{Z}_p^n}}\omega_p^{\frac{l}{2}(-(y+y')u'+2(z'-z)
  +w_1(x+x'))\times}\\[-4ex]
&&\hspace*{3cm} \sum_{v'}\omega_p^{\frac{l}{2}(v'(w_1-w_2+y'-y))}|u'+x-x',w_1\ra\la
u',w_2|,
\end{eqnarray*} where $u'=u-v$ and $v'=u+v$. Since $v'$ does
not occur in the quantum state, the sum over $v'$ vanishes unless
$w_2=w_1+y'-y$. Therefore, the state is \be
\sum_{\stackrel{(x,y,z),(x',y',z')\in H}{u',w_1\in
    \mathbb{Z}_p^n}}\omega_p^{\frac{l}{2}(-(y+y')u'+2(z'-z)
  +w_1(x+x'))}|u'+x-x',w_1\ra\la u',w_1+y'-y|. \nonumber \ee The
diagonal entries are obtained by putting $x=x'$ and $y=y'$ and since
$|H|=|S_H|$, we get $z=z'$. The diagonal entry is then proportional to
\be \sum_{\stackrel{(x,y,z)\in H}{u',w_1\in
    \mathbb{Z}_p^n}}\omega_p^{l(-yu'+w_1x)} .\nonumber \ee Up to
proportionality, this can be seen to be $\chi_{w_1,-u'}(H)$, a one
dimensional irrep. The bottom line is that, although not diagonal in
the Clebsch-Gordan basis, the resulting state's diagonal entries
correspond to one dimensional irreps we are interested in.

\section{The quantum algorithm}
\label{sec:qalgo}

In this section, we present a quantum algorithm that operates on two
copies of coset states at a time and show that it efficiently solves
the HSP over $G=\mathbb{Z}_p^{n+1}\rtimes\mathbb{Z}_p^n$, where the
input is $n$ and $\log p$. The algorithm is as follows:
\begin{enumerate}
\item Obtain two copies of coset states for $G$.
\item Perform a quantum Fourier transform on each of the coset states
  and measure the irrep label and row index for each state.  Assume
  that the measurement outcomes are high-dimensional irreps with
  labels $k$ and $l$. With high probability the irreps are indeed both
  high dimensional and $k+l\neq 0$, when $p>2$ (see the analysis
  below).  When $p=2$, there is only one high dimensional irrep which
  occurs with probability $1/2$ and $k+l=0$ always, since $k=l=1$. We
  deal with this case at the end of this section. For now assume that
  $p>2$ and $k+l\not=0$.
\item If $-k/l$ is not a square in $\Z_p$, then we discard the pair
  $(k,l)$ and obtain a new sample. Otherwise, perform a unitary
  $U_\alpha\otimes I:|u,v\ra\rightarrow |\alpha u,v\ra$, where
  $\alpha$ is determined by the two irrep labels as
  $\alpha=\sqrt{-k/l}$. This leads to a ``change" in the irrep
  label\footnote{We refer to Appendix \ref{ap:change} for a
    description of a technique that allows to change the labels of
    irreps of semidirect products that are more general than the
    Weyl-Heisenberg group.} of the first state from $k$ to $-l$. We
  can then apply the Clebsch-Gordan transform and obtain one
  dimensional irreps.
\item Apply a Clebsch-Gordan transform defined as \be
  U_{CG}:|u,v\ra\rightarrow \sum_{w\in \mathbb{Z}_p^n}
  \omega_p^{\frac{l}{2}(u+v)w}|u-v,w\ra \ee to these states.
\item Measure the two registers in the standard basis. With the
  measurement outcomes, we have to perform some classical
  post-processing which involves finding the orthogonal space of a
  vector space.
\end{enumerate} 

Now, we present the analysis of the algorithm.
\begin{enumerate}
\item In step 1, we prepared the state $\frac{1}{|G|}\sum_g|g\ra
  |0\ra$ and apply the black box $U_f$ to obtain the state
  $\frac{1}{|G|}\sum_g|g\ra |f(g)\ra$. After discarding the second
  register, the resulting state is $\frac{|H|}{|G|}|gH\ra\la gH|$.  We
  have two such copies.
\item After performing a QFT over $G$ on two such copies, we measure
  the irrep label and a row index. The probability of measuring an
  irrep label $\mu$ is given by $p(\mu)=d_\mu \chi_\mu(H)|H|/|G|$,
  where $\chi_\mu$ is the character of the irrep. If $\mu$ is a
  one-dimensional irrep, then the character is either $0$ or $1$ and
  so the probability becomes $0$ or $|H|/|G|$ accordingly. The character
  $\chi_\mu(H)=0$ if and only if $\mu=(a,b)\in S_H^\perp$. Therefore,
  the total probability of obtaining a one dimensional irrep is
  $|H||S_H^\perp|/|G|$. Now, we have that $|H|=|S_H|$ and so
  $|H||S_H^\perp|= p^{2n}$ since $S_H^\perp$ is the orthogonal space
  in $\mathbb{Z}_p^{2n}$.  Therefore, the total probability of
  obtaining a one dimensional irrep in the measurement is
  $p^{2n}/p^{2n+1}=1/p$. This is exponentially small in the input size
  ($\log p$). Therefore, the higher dimensional irreps occur with
  total probability of $1-1/p$.  Since all of them have the same
  $\chi_\mu(H)=p^n/|H|$, each of them occurs with the same probability
  of $1/p$. Take two copies of coset states and perform weak Fourier
  sampling and obtain two high dimensional irreps $k$ and $l$. The
  state is then $\frac{|H|^2}{p^{2n}}\rho_k(H)\otimes\rho_l(H)$. In
  the rest, we omit the normalization $\frac{|H|}{p^{n}}$ of each
  register.  Therefore, the state is proportional to \be
  \rho_k(H)\otimes\rho_l(H) = \sum_{(x,y,z),(x',y',z')\in H}
  \omega_p^{k(z+yu)+l(z'+y'v)}|u+x,v+y\ra\la u,v| .  \ee 
\item We can assume that $k$ and $l$ are such that $k+l\neq 0$ since
  this happens with probability $(p-1)/p^2$. Now, choose
  $\alpha=\sqrt{\frac{-k}{l}}$.  Since the equation $lx^2+k=0$ has at
  most two solutions for any $k,l\in \mathbb{Z}_p$, for any given
  $k,l$ chosen uniformly there exist solutions of the equation
  $lx^2+k=0$ with probability $1/2$.  Perform a unitary
  $U_\alpha:|u\ra\rightarrow |\alpha u\ra$ on the first copy. The
  first register becomes proportional to \ber
  U_\alpha \rho_k(H) U_\alpha^\dag &=& \sum_{(x,y,z)\in H} \omega_p^{k(z+yu)}|\alpha (u+x)\ra\la \alpha u| \nonumber \\
  &=& \sum_{(x,y,z)\in H,u_1\in\mathbb{Z}_p^n} \omega_p^{\frac{k}{\alpha^2}(z_1+y_1u_1)}|u_1+x_1\ra\la u| \nonumber \\
  &=& \rho_{\frac{k}{\alpha^2}}(\phi_\alpha(H)) , \eer where
  $(x_1,y_1,z_1)=\phi_\alpha(x,y,z)=(\alpha x, \alpha y, \alpha^2 z)$
  and $u_1=\alpha u$. It can be seen easily that $\phi_\alpha$ is an
  isomorphism of $G$ for $\alpha\neq 0$ and hence $\phi_\alpha(H)$ is
  subgroup of $G$. In fact, $\phi_\alpha(H)$ is a conjugate of $H$
  since $S_{\phi_\alpha(H)}=S_H$ (since if $(x,y)\in S_H$, then so is
  every multiple of it i.e., $(\alpha x,\alpha y)\in S_H$). Thus, we
  have obtained an irrep state with a new irrep label over a different
  subgroup. But this new subgroup is related to the old one by a known
  transformation. In choosing the value of $\alpha$ as above, we
  ensure that $k/\alpha^2=-l$ and hence obtain one dimensional irreps
  in the Clebsch-Gordan decomposition.
\item We now compute the state after performing a Clebsch-Gordan
  transform $U_{CG}$ on the two copies of the coset states, i.e.,
  perform the unitary given by the action \be
  U_{CG}:|u,v\ra\longrightarrow \sum_{w\in \mathbb{Z}_p^n}
  \omega_p^{\frac{l}{2}(u+v)w} |u-v,w\ra .  \ee The initial state of
  the two copies is \begin{eqnarray*}
    &&\rho_{-l}(\phi_\alpha(H))\otimes \rho_l(H)\\
    &=&\sum_{\stackrel{(x_1,y_1,z_1)\in\phi_\alpha(H),(x',y',z')\in
        H}{u,v\in\mathbb{Z}_p^n}} \omega_p^{-l(z_1+y_1u)+l(z'+y'v)}
    |u+x_1,v+x'\ra\la u,v|. \end{eqnarray*} The resulting state after
  the transform is
\begin{eqnarray*}
 && \hspace*{-1cm}\sum_{\stackrel{(x_1,y_1,z_1)\in\phi_\alpha(H),(x',y',z')\in H}{u,v,w_1,w_2\in\mathbb{Z}_p^n}} \omega_p^{-l(z_1+y_1u)+l(z'+y'v)+ \frac{l}{2}(u+v)(w_1-w_2)+(x_1+x')w_1}\times\\[-4ex] 
&&\hspace*{4cm} |u-v+x_1-x',w_1\ra\la u-v,w_2|\\[2ex]
 & =& \hspace*{-1cm}\sum_{\stackrel{(x_1,y_1,z_1)\in\phi_\alpha(H),(x',y',z')\in
      H}{u',v',w_1,w_2\in\mathbb{Z}_p^n}}
  \omega_p^{-l(z_1+y_1\frac{u'+v'}{2})+l(z'+y'\frac{v'-u'}{2})+
    \frac{l}{2}(v')(w_1-w_2)+(x_1+x')w_1}\times\\[-4ex]
&&\hspace*{4cm} |u'+x_1-x',w_1\ra\la u',w_2|,
\end{eqnarray*}
where $u'=u-v$ and $v'=u+v$. Notice that $v'$
  occurs only in the phase and not in the quantum states. Therefore,
  collecting the terms with $v'$ we get \be \sum_{v'}
  \omega_p^{\frac{l}{2}(y'-y_1+w_1-w_2)} .  \ee This term is non-zero
  only when $y'-y_1+w_1-w_2=0$. Hence $w_2=w_1-(y_1-y')$. Substituting
  this back in the equation, we get \begin{eqnarray*}
  &&\sum_{\stackrel{(x_1,y_1,z_1)\in \phi_\alpha(H),(x',y',z')\in
      H}{u',w_1\in\mathbb{Z}_p^n}}
  \omega_p^{\frac{l}{2}\left[(x_1+x')w_1-(y_1+y')u'-2(z_1-z')\right]}\\[-4ex]
  &&\hspace*{5cm}|u'+x_1-x',w_1\ra\la u',w_1-(y_1-y')|.\end{eqnarray*} Reusing the
  labels $u$ and $v$ by putting $u=u'$ and $v=w_1-(y_1-y')$, we obtain
  \begin{eqnarray*}
  &&\sum_{\stackrel{(x_1,y_1,z_1)\in \phi_\alpha(H),(x',y',z')\in
      H}{u,v\in\mathbb{Z}_p^n}}
  \omega_p^{\frac{l}{2}\left[(x_1+x')(v+(y_1-y'))-(y_1+y')u-2(z_1-z')\right]}\\[-4ex]
  &&\hspace*{5cm}|u+x_1-x',v+y_1-y'\ra\la u,v|.\end{eqnarray*} This can be written
  as \begin{eqnarray*} 
  &&\sum_{\stackrel{(x_1,y_1,z_1)\in \phi_\alpha(H),(x',y',z')\in
      H}{u,v\in\mathbb{Z}_p^n}}
  \omega_p^{\frac{l}{2}\left[(x_1+x')v-(y_1+y')u-2(z_1-\frac{x_1y_1}{2})+
  2(z'-\frac{x'y'}{2})\right]}\\[-4ex]  
  &&\hspace*{5cm}|u+x_1-x',v+y_1-y'\ra\la u,v|.\end{eqnarray*} Since $H$ is abelian,
  $x_1y'-x'y_1=0$. Now consider the subgroup $H_0$ defined in the
  previous section. Let $g=(\hat{x},\hat{y},\hat{z})$ be an element
  such that $H^g=H_0$. As discussed in Sec. \ref{sec:esg},
  $(\hat{x},\hat{y})$ are unique up to an element of $S_H^\perp$ and
  $\hat{z}$ is any element in $\mathbb{Z}_p$. Now, when $(x',y',z')\in
  H$ is conjugated with $g$, it gives
  $(x',y',z'+\hat{x}y'-\hat{y}x')=(x',y',x'y'/2)\in H_0$. Therefore,
  $z'-x'y'/2=x'\hat{y}-\hat{x}y'$. In order to obtain $H_0$ from
  $\phi_\alpha(H)$ we need to conjugate by
  $\phi_\alpha(\hat{x},\hat{y},\hat{z})$. Therefore,
  $z_1-\frac{x_1y_1}{2}=\alpha(\hat{y}x_1-\hat{x}y_1)$. Incorporating
  this into the above expression, we get 
  \begin{eqnarray*}
   && \hspace*{-1cm}\sum_{\stackrel{(x_1,y_1),(x',y')\in S_H}{u,v\in\mathbb{Z}_p^n}}
  \omega_p^{\frac{l}{2}\left[(x_1+x')v-(y_1+y')u-2(\alpha(\hat{y}x_1-\hat{x}y_1))+2(x'\hat{y}-\hat{x}y')\right]}\\[-4ex]
  &&\hspace*{2cm}|u+x_1-x',v+y_1-y'\ra\la u,v|.\end{eqnarray*} Now since $S_H$ is a
  linear space, we have that if $(x,y),(x',y') \in S_H$, then
  $(x-x',y-y') \in S_H$. Hence, substituting $x=x_1-x',y=y_1-y'$, we
  get 
  \begin{eqnarray*}
   && \hspace*{-1cm} \sum_{\stackrel{(x,y),(x',y')\in S_H}{u,v\in\mathbb{Z}_p^n}}
  \omega_p^{\frac{l}{2}\left[(x+2x')v-(y+2y')u-2(\alpha(\hat{y}(x+x')-\hat{x}(y+y')))+2(x'\hat{y}-\hat{x}y')\right]}\\[-4ex]
  &&\hspace*{2cm}|u+x,v+y\ra\la u,v|.\end{eqnarray*} Separating the sums over
  $(x,y)$ and $(x'y')$ we get 
  \begin{eqnarray*}
   && \hspace*{-1cm} \sum_{(x,y)\in
    S_H,u,v\in\mathbb{Z}_p^n} \left[\sum_{(x',y')\in
      S_H}\omega_p^{l\left[x'(v+(1-\alpha)\hat{y})-y'(u+(1-\alpha)\hat{x})\right]}
  \right]\\
  &&\hspace*{2cm}\omega_p^{\frac{l}{2}\left[x(v-2\alpha\hat{y})-y(u-2\alpha\hat{x})\right]}
  |u+x,v+y\ra\la u,v|.\end{eqnarray*} Note that the term in the
  squared brackets is non-zero only when
  $(v+(1-\alpha)\hat{y},u+(1-\alpha)\hat{x})$ lies in $S_H^\perp$.
  This means that if we measure the above state we obtain pairs
  $(u,v)$ such that $(u+(1-\alpha)\hat{x},v+(1-\alpha)\hat{y})\in
  S_H^\perp$. This can be used to determine both $S_H^\perp$ (and
  hence $S_H$) and $(\hat{x},\hat{y})$. Repeat this $O(n)$ times and
  obtain values for $u$ and $v$ by measurement.

\item From the above, say we obtain $n+1$ values
  $(u_1,v_1),\dots,(u_{n+1},v_{n+1})$. Therefore, we have the
  following vectors in $S_H^\perp$.  \ber
  (u_1+(1-\alpha_1)\hat{x},v_1+(1-\alpha_1)\hat{y}),\nonumber\\
  (u_2+(1-\alpha_2)\hat{x},v_2+(1-\alpha_2)\hat{y}), \nonumber \\
  \vdots \hspace{1.5in} \vdots \hspace{0.4in} \nonumber \\
  (u_{n+1}+(1-\alpha_{n+1})\hat{x},v_{n+1}+(1-\alpha_{n+1})\hat{y}) . \nonumber
  \eer The affine translation can be removed by first dividing by $(1-\alpha_i)$ and then taking the differences
  since $S_H^\perp$ is a linear space. Therefore, the following
  vectors lie in $S_H^\perp$: \ber
  (u_1',v_1')=(\frac{u_1}{(1-\alpha_1)}-\frac{u_{n+1}}{(1-\alpha_{n+1})},\frac{v_1}{(1-\alpha_1)}-\frac{v_{n+1}}{(1-\alpha_{n+1})}),\nonumber\\
  (u_2',v_2')=(\frac{u_2}{(1-\alpha_2)}-\frac{u_{n+1}}{(1-\alpha_{n+1})},\frac{v_2}{(1-\alpha_2)}-\frac{v_{n+1}}{(1-\alpha_{n+1})}), \nonumber \\
  \vdots \hspace{1.5in} \vdots \hspace{0.4in} \nonumber \\
  (u_n',v_n')=(\frac{u_n}{(1-\alpha_n)}-\frac{u_{n+1}}{(1-\alpha_{n+1})},\frac{v_n}{(1-\alpha_n)}-\frac{v_{n+1}}{(1-\alpha_{n+1})}) . \nonumber \eer With high
  probability, these vectors form a basis for $S_H^\perp$ and hence we
  can determine $S_H$ efficiently. This implies that the conjugacy
  class and hence the subgroup $H_0$ is known. It remains only to
  determine $(\hat{x},\hat{y})$. We can set
  $(\hat{x},\hat{y})=(1-\alpha_1)^{-1}(u_1-u'_1,v_1-v'_1)$ since the
  conjugating element can be determined up to addition by an element
  of $S_H^\perp$. $H$ can be obtained with the knowledge of $H_0$ and
  $(\hat{x},\hat{y})$.
\end{enumerate}

\noindent
Finally, for completeness we consider the case $p=2$. Assume that
after Fourier sampling we have two high dimensional irreps with states
given by \be \rho_1(H)\otimes\rho_1(H)=\sum_{(x,y,z),(x',y',z')\in
  H,u,v\in\mathbb{Z}_2^n}(-1)^{z+z'+yu+y'v}|u+x,v+x'\ra\la u,v|.  \ee
The Clebsch-Gordan transform is given by the base change: \be |u,v\ra
\rightarrow \sum_{w\in\mathbb{Z}_2^n}(-1)^{wv}|u+v,w\ra.  \ee Applying
this to the two states, we obtain (in a similar manner as above) \be
\sum_{(x,y,z)\in H,u,v\in\mathbb{Z}_2^n}(-1)^{z+vx}
\left(\sum_{(x',y',z')\in H}(-1)^{uy'+vx'}\right) |u+x,v+y\ra\la u,v|
.  \ee The inner sum is non-zero if and only if $(u,v)\in S_H^\perp$.
Thus, measuring this state gives us $S_H^\perp$ from which we can find
$S_H$. We cannot determine $H$ directly from here as in the case
$p>2$. But since we know $S_H$, we know the conjugacy class of $H$ and
we can determine the abelian group $HG'$ which contains $H$. This
group is obtained by appending the elements of $S_H$ with every
element of $G'=\mathbb{Z}_2$ i.e., for $(x,y)\in S_H$ we can say that
$(x,y,0)$ and $(x,y,1)$ are in $HG'$. Once we know $HG'$, we now
restrict the hiding function $f$ to the abelian subgroup $HG'$ of $G$
and run the abelian version of the standard algorithm to find $H$.
In summary, we have shown the following result: 
\begin{theorem}
  For $n {\geq} 1$, and $p{\geq} 2$ prime, the hidden subgroup problem for
  the Weyl-Heisenberg group $G$ of order $p^{2n+1}$ can be solved on a
  quantum computer with $O(n)$ queries. The time complexity of the
  quantum algorithm can be bounded by $O(n^3 \log p)$
  operations\footnote{Ignoring factors growing as $\log \log p$ or
    weaker.} and the algorithm uses at most $k=2$ coset states at the
  same time.
\end{theorem}

{\em Sketch of proof.} From the above discussion follows that $O(n)$
iterations of Steps 1.--4. in the algorithm will lead to system of
equations in Step 5. that with constant probability has a unique
solution. The number of queries in each iteration is constant and the
computational complexity of each of these steps can be upper bounded
as follows: $O(n \log p \, \log \log p)$ operations for each
computation of QFT over $G$ as described in Appendix \ref{ap:qft}. The
transform $U_\alpha$ and the Clebsch-Gordan transform $U_{CG}$ can
easily be implemented using arithmetic modulo $p$ and QFTs over
$\Z_p$, both of which can be done in $O(\log p\, \log \log p)$
elementary quantum operations.  Hence the running time of the quantum
part of the algorithm can be upper bounded by $O(n^2 \log p\, \log
\log p)$ operations and the number of queries by $O(n)$. The overall
running time is dominated by the cost for classical post-processing
which consists in computing the kernel of an $n\times n$ matrix over
$\Z_p$.  This can be upper bounded by $O(n^3)$ arithmetic operations
over $\Z_p$ for the Gaussian elimination, leading to a total bit
complexity of $O(n^3 \log p \, \log \log p \, 2^{O(\log^* \log p)})$
operations when using the currently fastest known algorithm for
integer multiplication \cite{Fuerer:2007}.  \hfill $\Box$

\section{Conclusions}\label{sec:conclusions}

Using the framework of coset states and non-abelian Fourier sampling
we showed that the hidden subgroup problem for the Weyl-Heisenberg
groups can be solved efficiently.  In each iteration of the algorithm
the quantum computer operates on $k=2$ coset states simultaneously
which is an improvement over the previously best known quantum
algorithm which required $k=4$ coset states. We believe that the
method of changing irrep labels and the technique of using
Clebsch-Gordan transforms to devise multiregister experiments has some
more potential for the solution of HSP over other groups.  Finally,
this group has importance in error correction.  In fact, the state we
obtain after Fourier sampling and measurement of an irrep is a
projector onto the code space whose stabilizer generators are given by
the generators of $H$. In view of this fact, it will be interesting to
study the implications of the quantum algorithm derived in this paper
to the design or decoding of quantum error-correcting codes.

\section*{Acknowledgments}
We thank Sean Hallgren and Pranab Sen for useful comments and discussions.


\newcommand{\etalchar}[1]{$^{#1}$}

\begin{appendix}
\section{QFT for the Weyl-Heisenberg groups}\label{ap:qft}

We briefly sketch how the quantum Fourier transform (QFT) can be
computed for the Weyl-Heisenberg groups
$G_n=\mathbb{Z}_p^{n+1}\rtimes\mathbb{Z}_p^n$. An implementation of
the QFT for the case where $p=2$ was given in \cite{Hoyer:97}. This
can be extended straightforwardly to $p>2$ as follows.
Using Eq.~(\ref{eq:QFTdef}), we obtain
that the QFT for $G_n$ is given by the unitary operator
\begin{eqnarray} {\rm QFT}_{G_n} &=&
  \sum_{a,b,x,y\in\mathbb{Z}_p^n, z\in\mathbb{Z}_p}
  \sqrt{\frac{1}{p^{2n+1}}}\omega_p^{ax+by}|0,a,b\ra\la z,x,y|\nonumber\\
  && \hspace*{1cm}+ \sum_{\stackrel{a,b,x,y\in\mathbb{Z}_p^n}{k\in\mathbb{Z}_p^\ast,z\in\mathbb{Z}_p}} \sqrt{\frac{p^n}{p^{2n+1}}}\omega_p^{k(z+by)}\delta_{x,a-b}|k,a,b\ra\la z,x,y| \nonumber \\
  &=& \sum_{\stackrel{a',b',x',y'\in\mathbb{Z}_p^{n-1}}{a_n,b_n,x_n,y_n,z\in\mathbb{Z}_p}} \sqrt{\frac{1}{p^{2n-1}}}\frac{1}{p}\omega_p^{a'x'+b'y'}\omega_p^{a_nx_n+b_ny_n}\nonumber \\[-3ex]
  && \hspace*{3cm} |0,a'a_n,b'b_n\ra\la z,x'x_n,y'y_n| \nonumber \\[2ex]
  &+&
  \sum_{\stackrel{k\in\mathbb{Z}_p^\ast,a',b',x',y'\in\mathbb{Z}_p^{n-1}}{z,a_n,b_n,x_n,y_n\in\mathbb{Z}_p}}\sqrt{\frac{p^{n-1}}{p^{2n-1}}}\frac{1}{\sqrt{p}}
  \omega_p^{k(z+b'y')}\omega_p^{ky_nb_n}\delta_{x',a'-b'}\delta_{x_n,a_n-b_n}
  \nonumber \\[-3ex]
  &&\hspace*{3cm} |k,a'a_n,b'b_n\ra\la z,x'x_n,y'y_n| \nonumber \\[2ex]
  &=& U\cdot {\rm QFT}_{G_{n-1}}\label{eq:rec}.  \end{eqnarray}

The matrix $U$ is given by \begin{eqnarray}\label{eq:U}
U &=& \sum_{x_n,y_n,a_n,b_n\in\mathbb{Z}_p} \frac{1}{p}\omega_p^{a_nx_n+b_ny_n} |0\ra\la 0| \otimes |a_n,b_n\ra\la x_n,y_n | \nonumber \\
&+& \sum_{x_n,y_n,a_n,b_n\in\mathbb{Z}_p,k\in\mathbb{Z}_p^\ast} \frac{1}{\sqrt{p}}\omega_p^{b_ny_n} \delta_{x_n,a_n-b_n}|k\ra\la k| \otimes |a_n,b_n\ra\la x_n,y_n | \nonumber \\
&=& |0\ra\la 0|\otimes {\rm QFT}_{\mathbb{Z}_p}\otimes {\rm
  QFT_{\mathbb{Z}_p}} + \sum_{k\in\mathbb{Z}_p^\ast} V\cdot
(I_p\otimes {\rm QFT}^{(k)}_{\mathbb{Z}_p}), \end{eqnarray} where $I_p$ is the
$p$ dimensional identity matrix, \begin{equation}
V=\sum_{u,v\in\mathbb{Z}_p}|u+v,v\ra\la u,v| , \ee and \be {\rm
  QFT}^{(k)}_{\mathbb{Z}_p}=\frac{1}{\sqrt{p}}\sum_{u,v\in\mathbb{Z}_p}\omega_p^{kuv}|u\ra\la
v|. \end{equation}  

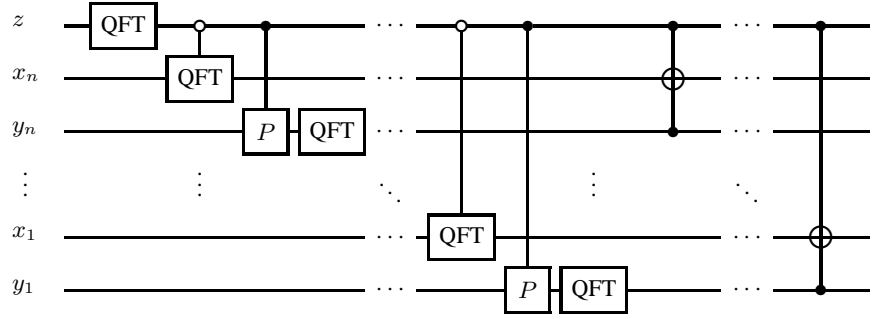
\begin{figure}[hbt]
\centerline{\unitlength1pt
\thicklines
\begin{picture}(20,120)
\put(0,0){\makebox{$y_1$}}
\put(0,20){\makebox{$x_1$}}
\multiput(5,36)(0,4){3}{\makebox(0,0){.}}
\put(0,60){\makebox{$y_n$}}
\put(0,80){\makebox{$x_n$}}
\put(0,100){\makebox{$z$}}
\end{picture}%
\begin{picture}(10,120)
\put(0,0){\line(1,0){10}}
\put(0,20){\line(1,0){10}}
\put(0,60){\line(1,0){10}}
\put(0,80){\line(1,0){10}}
\put(0,100){\line(1,0){10}}
\end{picture}%
\begin{picture}(24,120)
\put(0,92){\framebox(24,16){QFT}}
\put(0,0){\line(1,0){24}}
\put(0,20){\line(1,0){24}}
\put(0,60){\line(1,0){24}}
\put(0,80){\line(1,0){24}}
\end{picture}%
\begin{picture}(5,120)
\put(0,0){\line(1,0){5}}
\put(0,20){\line(1,0){5}}
\put(0,60){\line(1,0){5}}
\put(0,80){\line(1,0){5}}
\put(0,100){\line(1,0){5}}
\end{picture}%
\begin{picture}(24,120)
\put(0,72){\framebox(24,16){QFT}}
\multiput(12,36)(0,4){3}{\makebox(0,0){.}}
\put(0,0){\line(1,0){24}}
\put(0,20){\line(1,0){24}}
\put(0,60){\line(1,0){24}}
\put(0,100){\line(1,0){10}}
\put(14,100){\line(1,0){10}}
\put(12,100){\circle{4}}
\put(12,98){\line(0,-1){10}}
\end{picture}%
\begin{picture}(5,120)
\put(0,0){\line(1,0){5}}
\put(0,20){\line(1,0){5}}
\put(0,60){\line(1,0){5}}
\put(0,80){\line(1,0){5}}
\put(0,100){\line(1,0){5}}
\end{picture}%
\begin{picture}(16,120)
\put(0,52){\framebox(16,16){$P$}}
\put(0,0){\line(1,0){16}}
\put(0,20){\line(1,0){16}}
\put(0,80){\line(1,0){16}}
\put(0,100){\line(1,0){16}}
\put(10,100){\line(1,0){10}}
\put(8,100){\circle*{4}}
\put(8,98){\line(0,-1){30}}
\end{picture}%
\begin{picture}(5,120)
\put(0,0){\line(1,0){5}}
\put(0,20){\line(1,0){5}}
\put(0,60){\line(1,0){5}}
\put(0,80){\line(1,0){5}}
\put(0,100){\line(1,0){5}}
\end{picture}%
\begin{picture}(24,120)
\put(0,52){\framebox(24,16){QFT}}
\put(0,0){\line(1,0){24}}
\put(0,20){\line(1,0){24}}
\put(0,100){\line(1,0){24}}
\put(0,80){\line(1,0){24}}
\end{picture}%
\begin{picture}(20,120)
\multiput(10,0)(0,20){2}{\makebox(0,0){$\cdots$}}
\multiput(10,60)(0,20){3}{\makebox(0,0){$\cdots$}}
\put(10,40){\makebox(0,0){$\ddots$}}
\end{picture}%
\begin{picture}(5,120)
\put(0,0){\line(1,0){5}}
\put(0,20){\line(1,0){5}}
\put(0,60){\line(1,0){5}}
\put(0,80){\line(1,0){5}}
\put(0,100){\line(1,0){5}}
\end{picture}%
\begin{picture}(24,120)
\put(0,12){\framebox(24,16){QFT}}
\put(0,0){\line(1,0){24}}
\put(0,80){\line(1,0){24}}
\put(0,60){\line(1,0){24}}
\put(0,100){\line(1,0){10}}
\put(14,100){\line(1,0){10}}
\put(12,100){\circle{4}}
\put(12,98){\line(0,-1){70}}
\end{picture}%
\begin{picture}(5,120)
\put(0,0){\line(1,0){5}}
\put(0,20){\line(1,0){5}}
\put(0,60){\line(1,0){5}}
\put(0,80){\line(1,0){5}}
\put(0,100){\line(1,0){5}}
\end{picture}%
\begin{picture}(16,120)
\put(0,-8){\framebox(16,16){$P$}}
\put(0,20){\line(1,0){16}}
\put(0,60){\line(1,0){16}}
\put(0,80){\line(1,0){16}}
\put(0,100){\line(1,0){6}}
\put(10,100){\line(1,0){6}}
\put(8,100){\circle*{4}}
\put(8,98){\line(0,-1){90}}
\end{picture}%
\begin{picture}(5,120)
\put(0,0){\line(1,0){5}}
\put(0,20){\line(1,0){5}}
\put(0,60){\line(1,0){5}}
\put(0,80){\line(1,0){5}}
\put(0,100){\line(1,0){5}}
\end{picture}%
\begin{picture}(24,120)
\put(0,-8){\framebox(24,16){QFT}}
\multiput(12,36)(0,4){3}{\makebox(0,0){.}}
\put(0,100){\line(1,0){24}}
\put(0,20){\line(1,0){24}}
\put(0,60){\line(1,0){24}}
\put(0,80){\line(1,0){24}}
\end{picture}%
\begin{picture}(10,120)
\put(0,0){\line(1,0){10}}
\put(0,20){\line(1,0){10}}
\put(0,60){\line(1,0){10}}
\put(0,80){\line(1,0){10}}
\put(0,100){\line(1,0){10}}
\end{picture}%
\begin{picture}(16,120)
\put(0,0){\line(1,0){16}}
\put(8,80){\circle{8}}
\put(8,60){\circle*{4}}
\put(8,100){\circle*{4}}
\put(8,100){\line(0,-1){40}}
\put(0,20){\line(1,0){16}}
\put(0,60){\line(1,0){16}}
\put(0,80){\line(1,0){16}}
\put(0,100){\line(1,0){16}}
\end{picture}%
\begin{picture}(10,120)
\put(0,0){\line(1,0){10}}
\put(0,20){\line(1,0){10}}
\put(0,60){\line(1,0){10}}
\put(0,80){\line(1,0){10}}
\put(0,100){\line(1,0){10}}
\end{picture}%
\begin{picture}(20,120)
\multiput(10,0)(0,20){2}{\makebox(0,0){$\cdots$}}
\multiput(10,60)(0,20){3}{\makebox(0,0){$\cdots$}}
\put(10,40){\makebox(0,0){$\ddots$}}
\end{picture}%
\begin{picture}(10,120)
\put(0,0){\line(1,0){10}}
\put(0,20){\line(1,0){10}}
\put(0,60){\line(1,0){10}}
\put(0,80){\line(1,0){10}}
\put(0,100){\line(1,0){10}}
\end{picture}%
\begin{picture}(16,120)
\put(0,0){\line(1,0){16}}
\put(8,20){\circle{8}}
\put(8,0){\circle*{4}}
\put(8,100){\circle*{4}}
\put(8,100){\line(0,-1){100}}
\put(0,20){\line(1,0){16}}
\put(0,60){\line(1,0){16}}
\put(0,80){\line(1,0){16}}
\put(0,100){\line(1,0){16}}
\end{picture}%
\begin{picture}(10,120)
\put(0,0){\line(1,0){10}}
\put(0,20){\line(1,0){10}}
\put(0,60){\line(1,0){10}}
\put(0,80){\line(1,0){10}}
\put(0,100){\line(1,0){10}}
\end{picture}%
}
\vspace*{0.3cm}
\caption{\label{fig:qft} QFT for the Weyl-Heisenberg group. The QFT
  gates shown in the circuit are QFTs for the cyclic groups
  $\mathbb{Z}_p$.  Each of these QFTs can be implemented approximately
  \cite{Kitaev:97,HH:2000} or exactly \cite{MZ:2004}, in both cases
  with a complexity bounded by $O(\log p\, \log \log p)$. It should be
  noted that the wires in this circuit are actually $p$-dimensional
  systems.  The meaning of the controlled gates where the control wire
  is an open circle is that the operation is applied to the target
  wire if and only if the control wire is in the state $\ket{0}$. The
  meaning of the controlled $P$ gates where the control wire is a
  closed circle here means that the gate $P_k$ is applied in case the
  control wire is in state $\ket{k}$ with $k\not=0$, and $P_0=I_p$.
  Here $P_k$ is the permutation matrix for which ${\rm QFT}^{(k)}=P_k
  {\rm QFT}$ holds.  The complexity of this circuit can be bounded by
  $O(n \log p\, \log \log p)$.}
\end{figure}

From Eq.~(\ref{eq:U}) and recursive application of Eq.~(\ref{eq:rec})
we obtain the efficient quantum circuit implementing ${\rm QFT}_{G_n}$
shown in Figure \ref{fig:qft}.

\section{Changing labels of irreducible representations}
\label{ap:change}

In this section, we describe the technique of changing labels of
irreducible representations (irreps) in a more abstract,
representation theoretic, fashion. We consider a situation slightly
more general than the Weyl-Heisenberg groups considered in the paper,
namely for semidirect products of the form $G=A\rtimes_\phi B$, where
$A$ is an Abelian group, $B$ is an arbitrary finite group, and
$\phi:B\rightarrow \rm{Aut}(A)$. We make some further assumptions
regarding the irreps of $G$ that arise during Fourier sampling.
First, note that in general there might be some irreps of $G$ that
arise as inductions \cite{Serre:77,Huppert:83} of irreps of $A$ to
$G$. Suppose that, with high probability, we sample only such irreps,
so that we can restrict our attention to this case. This happens for
the Weyl-Heisenberg groups discussed in this paper.  Other examples
are the groups isomorphic to $\mathbb{Z}_p^n\rtimes\mathbb{Z}_p$
studied in \cite{BCvD:2005} and the affine groups \cite{MRRS:2004}
which are isomorphic to $\mathbb{Z}_p\rtimes\mathbb{Z}_{p-1}$.

After Fourier sampling and measurement of an irrep label we have the
state $\rho_k(H)$, where $\rho_k$ is an irrep of $G$ and $k$ is its
label. We want to apply an operator $U_B$ to this state in order to
change it to a state $\rho_{k'}(H')$ corresponding to an irrep with
label $k'$, possibly with respect to a different subgroup $H'$.  In
the following we show how this can be done if $\rho_k(H) =
(\chi_k\uparrow G)(H)$, i.\,e., if $\rho_k$ is an induction of an
irrep $\chi_k$ of $A$ to $G$. The possible labels $k'$ that can be
obtained depend on the automorphism group of $B$, namely on those
automorphisms of $B$ that can be extended to automorphisms of $G$.

First, recall that for $\chi_k \in \hat{A}$, the image of an element
$(a,b)\in G$ under the induction of $\chi_k$ to $G$ is given by
\begin{equation}\label{eq:ind}
(\chi_k\uparrow G)(a,b) = \sum_{t\in B} \chi_k(\phi_{t^{-1}}(a))|t
b^{-1}\ra\la t|,
\end{equation}
where $\phi_{t^{-1}}=(a \mapsto \phi^{-1}(t)(a))\in \rm{Aut}(A)$.  Now
consider an automorphism of $B$, say $\beta\in {\rm Aut}(B)$.  Let
$U_B$ be the unitary matrix acting on $\mathbb{C}[B]$ corresponding to this
automorphism. Applying $U_B$ to Eq.~(\ref{eq:ind}), we get 
\begin{equation}\label{eq:step1}
\sum_{t\in B}\chi_k(\phi_{t^{-1}}(a))|\beta(t) \beta(b^{-1})\ra\la
\beta(t)| = \sum_{t\in B} \chi_k(\phi_{\beta(t)}(a))|t
\beta(b^{-1})\ra\la t| .
\end{equation}
In order to further simplify this expression, we now suppose that we
can extend the automorphism $\beta$ to an automorphism of the whole
group in the form $\gamma=(\alpha,\beta)\in {\rm Aut}(G)$, where
$\alpha\in\rm{Aut}(A)$. We derive some conditions that $\alpha$ has to
satisfy in order for this extension to be possible. First, we have
that \be \gamma ((a_1,b_1)(a_2,b_2))=\gamma(a_1,b_1) \gamma(a_2,b_2).
\ee This condition becomes 
\begin{equation}\label{eq:aut}
((\alpha\phi_{b_2})(a_1) +
\alpha(a_2),b_1b_2)=((\phi_{\beta(b_2)}\alpha)(a_1) +
\alpha(a_2),\beta(b_1b_2)).
\end{equation}
Note that in the above equation, since $\alpha$ and $\phi_t$ are
elements of $\rm{Aut}(A)$ for all $t$, we write their product acting
on $a\in A$ as $(\alpha\phi_t)(a)$.  From Eq.~(\ref{eq:aut}) we obtain
that \be\label{Cond} \phi_{\beta(b)} = \alpha\phi_{b}\alpha^{-1} \ee
for all $b\in B$. This means that $\alpha\in
N_{\rm{Aut}(A)}(\rm{Im}(\phi))$ i.e., $\alpha$ lies in the normalizer of
$\rm{Im}(\phi)$, the image of $\phi$ in $\rm{Aut}(A)$.  Therefore, we
need to pick the pair $(\alpha,\beta)$ such that the condition in
Eq.~(\ref{Cond}) holds. It is clear that given $\alpha$ there always
exists $\beta$ such that Eq.~(\ref{Cond}) holds but not necessarily
the other way around.

Thus, using the assumption that the automorphism can be extended to
all of $G$, we can rewrite Eq.~(\ref{eq:step1}) as follows: \be
\sum_{t\in B} \chi_k(\phi_{\beta(t)}(a))|t \beta(b^{-1})\ra\la t| =
\sum_{t\in B} \chi_k((\alpha^{-1}\phi_{t^{-1}}\alpha) (a))|t
\beta(b^{-1})\ra\la t| .  \ee Now, the inner product
$\chi_k((\alpha^{-1}\phi_{t^{-1}}\alpha )(a))$ can be written as
$\chi_{\hat{\alpha}^{-1} k}((\phi_{t^{-1}}\alpha) (a))$. Therefore,
the state is given by \be \sum_{t\in B} \chi_{\hat{\alpha}^{-1}
  k}(\phi_{t^{-1}}(\alpha (a)) |t \beta(b)^{-1}\ra\la t| =
(\chi_{k'}\uparrow G)(\gamma(a,b)) , \ee where
$k'=\hat{\alpha}^{-1}(k)$. Here, $\hat{\alpha}$ is an automorphism of
the dual group $\hat{A}$ corresponding to $\alpha$ such that the
character remains invariant.  Overall, we have shown the following:
\begin{theorem}
  Let $G=A\rtimes_\phi B$ and $\rho_k = (\chi_k \uparrow G) \in
  \hat{G}$, where $\chi_k \in \hat{A}$. Let $U_B\in \C[B]$ be the
  unitary matrix corresponding to an automorphism $\beta\in {\rm
    Aut}(B)$ that can be extended to $\gamma = (\alpha,\beta)\in
  {\rm Aut}(G)$. Then by applying $U_B$ to the hidden subgroup state
  $\rho_k$, we can change it to: \be U_B\rho_k(H)U_B^\dag =
  \rho_{k'}(\gamma(H)),  \ee
where $k'=\hat{\alpha}^{-1}(k)$.
\end{theorem}


\end{appendix}

\end{document}